\documentclass[pra,twocolumn,graphicx]{revtex4}
\usepackage{epsfig,amsmath}

\begin{document}

\title{Classical and quantum-mechanical treatments of nonsequential double
ionization with few-cycle laser pulses}
\author{C. Figueira de Morisson Faria$^1$, X. Liu$^2$, A. Sanpera$^1$ and M.
Lewenstein$^1$}
\affiliation{$^1$Insitut f\"ur theoretische Physik, Universit\"at Hannover, Appelstr. 2,
30167 Hannover, Germany\\
$^2$Max-Born-Institut, Max-Born-Str. 2A, 12489 Berlin, Germany}
\date{\today}

\begin{abstract}
We address nonsequential double ionization induced by strong, linearly
polarized  laser fields of only a few cycles, considering a physical 
mechanism in
which the second electron is dislodged by the inelastic collision of the
first electron with its parent ion. The problem is treated classically,
using an ensemble model, and quantum-mechanically, within the
strong-field and uniform saddle-point approximations. In the latter case,
the results are interpreted in terms of ``quantum orbits", which can be
related to the trajectories of a classical electron in an electric field. We
obtain highly asymmetric electron momentum distributions, which strongly
depend on the absolute phase, i.e., on the phase difference between the
pulse envelope and its carrier frequency. Around a particular value of this
parameter, the distributions shift from the region of positive to that of 
negative momenta, or vice-versa, in a radical fashion. This
behavior is investigated in detail for several driving-field parameters, and
provides a very efficient method for measuring the absolute phase. Both
models yield very similar distributions, which share the same
physical explanation. There exist, however, minor discrepancies due to the 
fact that, beyond the 
region for which electron-impact ionization is classically allowed, the 
yields from the quantum mechanical computation decay exponentially, whereas 
their classical counterparts vanish.
\end{abstract}

\maketitle

\section{Introduction}

Linearly polarized laser pulses of intensities higher than $10^{14}\mathrm{%
W/cm}^{2}$ and only a few cycles are of vital importance to several areas of
physics, being applicable to, for instance, solid-state physics \cite{damage}%
, high-frequency sources \cite{fewcreview}, or isolated attosecond pulses 
\cite{atto1}. Only the latter application led to a breakthrough in
metrology, making it possible to trace the motion of bound electrons \cite%
{metrology}, to probe molecular dynamics \cite{attomol}, and to control
electron emission \cite{attocontrol}. In this pulse-length regime, the phase
difference between the pulse envelope and its carrier frequency, known as
``absolute phase'', has a major influence
on strong-field optical phenomena, such as high-order harmonic generation
(HHG) \cite{fewcyclehhg}, or above-threshold ionization (ATI) \cite{fewcati1}. 
In particular, the absolute phase affects, for instance, the harmonic or 
photoelectron yields, the maximal energies in both spectra and the time 
profiles of ATI and HHG. This is a direct consequence of the physical 
mechanisms governing such phenomena, which occur in a subfemtosecond 
time scale, 
and for which the time dependence of the electric field is important. In 
fact, high-order harmonic
generation is the outcome of a three-step process in which an electron
leaves an atom by tunneling ionization at a time$\ t^{\prime }$, propagates
in the continuum, and recombines with its parent ion at a later instant $t$,
releasing the energy gained from the field in form of high-frequency
radiation. A similar mechanism is also responsible for ATI, with the main
difference that the electron either rescatters elastically with its parent
ion, or reaches the detector without recolliding, originating high- or
low-energy peaks in the spectra, respectively. Such a mechanism has been
extensively studied both classically \cite{corkum,aticlass} and
quantum-mechanically \cite{tstep}.

From the experimental point of view, controlling or measuring the absolute
phase is a very difficult task \cite{phasecontrol}. This has led to the
proposal and experimental realization of schemes for its diagnosis, as, for
instance, using the asymmetry in ATI photoelectron counts reaching two
opposite detectors placed in a plane perpendicular to the laser field \cite%
{fewcati2}.

Another phenomenon whose physical explanation lies on a laser-assisted
rescattering process is nonsequential double ionization (NSDI). In this
case, an electron recollides inelastically with its parent ion, giving part
of its kinetic energy to a second electron, which is thus able to overcome
the second ionization potential and reach the continuum. Fingerprints of
such a mechanism were only revealed very recently, in experiments in which
the momentum component parallel to the laser field polarization could be
resolved, either for the doubly charged ion \cite{expe1ion}, or for both 
electrons \cite{expe1}. Such features, namely a
doubly-peaked structure in the momentum distributions, with maxima at $%
p_{1\parallel }=p_{2\parallel }=\pm 2\sqrt{U_{p}},$ where $p_{j\parallel
}(j=1,2)$ and $U_{p}$ denote the electron momentum components parallel to the laser
field polarization and the ponderomotive energy \cite{footnoteup}, respectively, are, up to
the present date, the most striking example of electron-electron correlation
in the context of atoms in strong laser fields. This fact has led not only
to further experiments \cite{moreexp}, but also to a considerable
theoretical activity on the subject, using quantum mechanical 
\cite{tdse,abeckerrep,abecker,richard,HLS}, semi-classical 
\cite{doublegoresl,nsdiuni,preprint1,preprint2} and classical 
\cite{class,slowdown,fewcycleclass} methods.

Recently, we have shown that NSDI may serve as a powerful tool
for absolute-phase measurements exploiting the fact that, for few-cycle
driving pulses, inversion symmetry is broken \cite{fewcycleclass}. Thus, the
distributions in $\left( p_{1\parallel },p_{2\parallel }\right) $ are mainly
concentrated in the positive or negative momentum regions, changing from one 
region to the other upon a critical phase. Such investigations have been 
performed classically, considering  electrons released at times 
$t^{\prime }$ uniformly distributed throughout the pulse and weighted with the
quasi-static tunneling rate \cite{quasistatic}.

In this paper, we deal with this problem quantum-mechanically, and investigate 
the existence of a one-to-one correspondence with the classical model in 
Ref. \cite{fewcycleclass}. 
Similar studies have been performed for NSDI with monochromatic driving 
fields, with practically identical outcomes \cite{preprint1,preprint2}. 
This has shown that, at least in the monochromatic case, which is a good 
approximation for the long pulses used in the experiments, intrinsically quantum-mechanical effects 
such as interference processes, or wave-packet spreading, are not important. 
However, it is legitimate to ask the question of whether this situation 
will persist in the few-cycle regime. Indeed, it may well be that 
interference and wave-packet spreading play a more important role in 
this latter case. Additionally, it is not clear whether the quasi-static 
tunneling rate considered in the classical model remains valid for few-cycle 
driving pulses. In fact, this has been recently put into question, 
with the derivation of a non-adiabatic rate \cite{rateyudivanov}. 
Finally, it is worth checking whether asymmetric distributions and the 
critical phase also occur in a quantum-mechanical context, and, 
in case they do, to understand the physics behind.

 In particular, we address the above-stated questions using a
$S$-Matrix formalism, within the Strong-Field Approximation (SFA) 
\cite{footnsfa,kfr}. We consider the
simplest type of rescattering, namely electron-impact ionization, and treat
the problem in terms of the so-called ``quantum orbits'' \cite{orbits}, which appear in the context of
saddle-point approximations. Specifically, we use a uniform approximation
whose only validity requirement is that the orbits in question occur in
pairs, which is in general the case for laser-assisted rescattering
phenomena. This method has been previously applied to NSDI in monochromatic
driving fields, in order to analyze the influence of the types of interaction
and final-state electron correlation on the yields 
\cite{preprint1,preprint2}. Apart from considerably simplifying the 
computations involved, as compared to
other theoretical methods \cite{tdse,abeckerrep,abecker}, the quantum-orbit approach
provides additional physical insight, in terms of a space-time picture. In
fact, the quantum orbits are closely related to the orbits of a classical
electron in an external laser field. Hence, in several situations, it is
possible to draw a parallel betweeen our quantum-mechanical treatment and
the previous classical considerations \cite{fewcycleclass}, discussing their
similarities and differences. In the following, we study the physical
mechanisms responsible for the critical phase within a quantum-mechanical
framework, concentrating on the main differences from the classical picture
and from the monochromatic-driving field case.

The paper is organized as follows: In the next section (Sec. \ref{backgrd}), we
provide the necessary theoretical background, presenting the transition
amplitude in the strong-field and uniform approximations. Subsequently, in
Sec. \ref{quantumorb}, we present differential electron momentum
distributions for various absolute phases, discussing the main features
obtained in terms of quantum orbits. The quantum mechanical results are then
compared to a classical ensemble computation which is either the same as in 
\cite{fewcycleclass}, or slightly modified with respect to it (Sec. \ref%
{classical}). Finally, in Sec. \ref{concl}, we summarize the paper and state
our conclusions.

\section{Background}

\label{backgrd}

The transition amplitude of the laser-assisted inelastic rescattering
process responsible for NSDI, in the Strong-Field Approximation \cite%
{footnsfa,kfr}, is given by 
\begin{eqnarray}
M&=&-\int_{-\infty }^{\infty }dt\int_{-\infty }^{t}dt^{\prime }\left\langle
\psi _{p_{1},p_{2}}^{(V)}(t)\right\vert V_{12}U_{1}^{(V)}(t,t^{\prime
})V  \notag \\
 &&{}\otimes U_{2}^{(0)}(t,t^{\prime
})\left\vert \psi _{0}(t^{\prime })\right\rangle,  \label{rescatt}
\end{eqnarray}%
where $V$, $U_{n}^{(0)}(t,t^{\prime })$, $U_{n}^{(V)}(t,t^{\prime }),$ and $V_{12}$ denote the atomic
binding potential, the field-free and the 
Volkov time evolution operators acting on the $n$-th $
(n=1,2)$ electron, and the interaction through which the second electron is
freed by the first, respectively. Eq. (\ref{rescatt}) expresses the
following physical process: Initially, both electrons are bound, and the
atom is in the ground state, which is approximated by $\left\vert \psi
_{0}(t^{\prime })\right\rangle =\left\vert \psi _{0}^{(1)}(t^{\prime
})\right\rangle \otimes \left\vert \psi _{0}^{(2)}(t^{\prime })\right\rangle 
$(i.e., product state of one-electron ground states), with $\left\vert \psi
_{0}^{(n)}(t^{\prime })\right\rangle =e^{i|E_{0n}|t^{\prime }}\left\vert
\psi _{0}^{(n)}\right\rangle .$ At the time $t^{\prime }$, the first
electron is released through tunneling ionization, whereas the second
electron remains bound. Subsequently, the first electron propagates in the
continuum from $t^{\prime }$ to $t$, gaining energy from the field. At this
latter time, it collides inelastically with its parent ion, dislodging the
second electron$.$ The final electron state is then chosen as the product
state of one-electron Volkov waves, $\left\vert \psi
_{p_{1},p_{2}}^{(V)}(t)\right\rangle =$ $\left\vert \psi
_{p_{1}}^{(V)}(t)\right\rangle $ $\otimes \left\vert \psi
_{p_{2}}^{(V)}(t)\right\rangle $, where $\mathbf{p_{1}},\mathbf{p_{2}}$ are
the final electron momenta (for studies of correlated two-electron final
states see, e.g., Refs. \cite{abeckerrep}, \cite{preprint1} and 
\cite{preprint2}). In Eq. (\ref{rescatt}), the interaction with the ionic 
potential is not taken into account. In our computations, we use the length 
gauge and atomic units.

Expanding $U^{(V)}(t,t^{\prime })$ in terms of Volkov states, Eq. 
(\ref{rescatt}) reads 
\begin{equation}
M=-\int_{-\infty }^{\infty }dt\int_{-\infty }^{t}dt^{\prime }\int d^{3}kV_{%
\mathbf{pk}}V_{\mathbf{k}0}\exp [iS(t,t^{\prime },\mathbf{p}_{i},\mathbf{k}%
)],  \label{prescatt}
\end{equation}
with the action 
\begin{eqnarray}
S(t,t^{\prime },\mathbf{p}_{i},\mathbf{k})&=&-\frac{1}{2}
\sum_{i=1}^{2}\int_{t}^{\infty }[\mathbf{p}_{i}+\mathbf{A}(\tau )]^{2}d\tau \notag \\
&&{}-\frac{1}{2}\int_{t^{\prime }}^{t}[\mathbf{k}+\mathbf{A}(\tau )]^{2}d\tau \notag \\
&&{}+|E_{01}|t^{\prime }+|E_{02}|t,  \label{action}
\end{eqnarray}%
where $\mathbf{A}(t),$ $\mathbf{p}_{j}(j=1,2),\mathbf{k}$ and $|E_{0n}|$ 
denote the vector potential, the final momenta of both electrons, the 
intermediate momentum of the first electron and the
ionization potentials, respectively. All the influence of the binding
potential $V$ and of the electron-electron interaction $V_{12}$ is included
in the form factors 
\begin{equation}
V_{\mathbf{pk}}=<\mathbf{p}_{2}+\mathbf{A}(t),\mathbf{p}_{1}+%
\mathbf{A}(t)| V_{12}| \mathbf{k}+\mathbf{A}(t),\psi
_{0}^{(2)}>
\end{equation}%
and 
\begin{equation}
V_{\mathbf{k}0}= <\mathbf{k}+\mathbf{A}(t^{\prime })|
V| \psi _{0}^{(1)}> .
\end{equation}%
In this paper, we consider a contact-type interaction 
\begin{equation}
V_{12}=\delta (\mathbf{r}_{1}-\mathbf{r}_{2})\delta (\mathbf{r}_{2}),
\label{contact}
\end{equation}%
which yields very good agreement with experimental data within the context
of NSDI in monochromatic driving fields \cite{doublegoresl,preprint1}. In
this case, the form factors $\ V_{\mathbf{pk}},V_{\mathbf{k}0}$ are constant
and the SFA transition amplitude can be solved analytically up to one
quadrature. For other types of potentials, this is only possible by evaluating
multiple integrals numerically. 

For low enough frequencies and high enough laser intensities, Eq. (\ref%
{prescatt}) can be solved to a good approximation by the steepest descent
method. Thus, we must determine $\mathbf{k}$, $t^{\prime }$ and $t$, such
that $S(t,t^{\prime },\mathbf{p}_{i},\mathbf{k})$ is stationary, i.e., 
 its partial derivatives with respect to these parameters vanish. This
condition yields 
\begin{equation}
\left[ \mathbf{k}+\mathbf{A}(t^{\prime })\right] ^{2}=-2|E_{01}|,
\label{saddle1}
\end{equation}
\begin{equation}
\sum_{j=1}^{2}\left[ \mathbf{p}_{j}+\mathbf{A}(t)\right] ^{2}=\left[ \mathbf{%
k}+\mathbf{A}(t)\right] ^{2}-2|E_{02}|  \label{saddle2}
\end{equation}
and
\begin{equation}
\int_{t^{\prime }}^{t}d\tau \left[ \mathbf{k}+\mathbf{A}(\tau )\right] =0.
\label{saddle3}
\end{equation}%
Eq. (\ref{saddle1}) and (\ref{saddle2}) give the energy conservation at the
start and rescattering times, respectively, while Eq. (\ref{saddle3})
constraints the intermediate momentum of the first electron so that it
returns to its parent ion. For vanishing $|E_{01}|,$ the classical equations
of motion of both electrons in the external field are obtained. For non-zero
$|E_{01}|$, Eq. (\ref%
{saddle1}) expresses tunneling ionization at $t^{\prime },$ and has no real
solution. Physically, this means that this process is not classically
allowed. This results in complex variables $t^{\prime },t$ and $\mathbf{k,}$
which always occur in pairs. The real parts of such variables are directly
related to a longer and a shorter orbit of a classical electron in an
electric field. The longer orbit can be associated to the so-called
\textquotedblleft slow-down collisions\textquotedblright , which have
recently been discussed in the literature \cite{HLS,nsdiuni,slowdown}. The
imaginary parts determine to which extent electron-impact ionization is
allowed or forbidden, both within and beyond the boundaries of the 
classically allowed energy region. In this latter domain, one of the 
orbits leads to exponentially decaying contributions in the 
transition amplitude (\ref{prescatt}), which cause 
cutoffs in the distributions, while the remaining orbit starts to
yield diverging contributions, and must be discarded.

Eq. (\ref{saddle2}) can also be written in terms of the momentum components
parallel and perpendicular to the laser-field polarization, denoted by $%
p_{j\parallel }$ and $\mathbf{p}_{j\perp }(j=1,2)$, respectively. In this
case, for constant transverse momenta, one obtains the equation 
\begin{equation}
\sum\limits_{j=1}^{2}[p_{j\parallel }+A(t)]^{2}=[\mathbf{k}+\mathbf{A}%
(t)]^{2}-2|E_{02}|-\sum\limits_{j=1}^{2}\mathbf{p}_{j\perp }^{2},
\label{saddle4}
\end{equation}%
describing a circle in the $p_{1\parallel },$ $p_{2\parallel }$ plane, whose 
radius depends on the kinetic energy 
$E_{\mathrm{ret}}(t)=[\mathbf{k}+\mathbf{A}(t)]^{2}/2$ of the first electron 
upon return, and on the effective binding
energy $|\tilde{E}_{02}|=|E_{02}|+\sum\limits_{j=1}^{2}\mathbf{p}_{j\perp
}^{2}/2.$ If $E_{\mathrm{ret}}(t)\leq |\tilde{E}_{02}|$, this radius
collapses and electron-impact ionization becomes classically forbidden. This
means that there is not only a maximal, but also a minimal classically
allowed energy, and the resulting yields exhibit two cutoffs, or no cutoff
at all. This is a major difference with respect to high-order harmonic
generation or above-threshold ionization, for which only maximal classically
allowed energies exist.

In the standard saddle point method, the action (\ref{action}) is expanded
quadratically around the saddle points, and the transition amplitude (\ref%
{prescatt}) is approximated by 
\label{sadresc}
\begin{eqnarray}
M^{(\mathrm{SPA})} &=&\sum_{s}A_{s}\exp (iS_{s}),  \label{sadresca} \\
S_{s} &=&S_{\mathbf{p}}(t_{s},t_{s}^{\prime },\mathbf{k}_{s}), \\
A_{s} &=&(2\pi i)^{5/2}\frac{V_{\mathbf{p}\mathbf{k}_{s}}V_{\mathbf{k}_{s}0}%
}{\sqrt{\det S_{\mathbf{p}}^{\prime \prime }(t,t^{\prime },\mathbf{k})|_{s}}}%
,  \label{sadrescc}
\end{eqnarray}%
where the index $s$ runs over the relevant saddle points, and $S_{\mathbf{p}%
}^{\prime \prime }$ denotes the five-dimensional matrix of the second
derivatives of the action with respect to $t,t^{\prime }$ and $\mathbf{k}$.
In practice, we first determine $\mathbf{k}(t,t^{\prime })$ as a function of
the other variables, inserting this in the action, and take 
\begin{equation}
A_{s}=(2\pi i)^{5/2}\frac{V_{\mathbf{p}\mathbf{k}_{s}}V_{\mathbf{k}_{s}0}}{%
(t^{\prime }-t)^{3/2}\sqrt{\det S_{\mathbf{p}}^{\prime \prime }(t,t^{\prime
})|_{s}}},  \label{detred}
\end{equation}%
so that the computation of the determinant is simplified.

The above-stated saddle-point approximation is only applicable for
well-isolated saddle-points. This does not hold near the boundaries of the
classically allowed region, where the pairs of saddles nearly coalesce.
Furthermore, beyond such boundaries, one of the saddles yields diverging
results, and must be discarded. This leads to cusps in the yield which are
particularly problematic for non-sequential double ionization. A detailed
analysis of this problem is given in \cite{nsdiuni}.

Such artifacts can be eliminated by using a more general, uniform
approximation \cite{Bleistein}, whose only validity requirement is that the
saddles occur in pairs. This approximation
 has been successfully applied in the context of above-threshold ionization 
\cite{atiuni} and nonsequential double ionization 
\cite{nsdiuni,preprint2}.

Within this improved approximation, in the classically allowed region, the
transition amplitude for a pair of trajectories $i$ and $j$ is given by 
\begin{eqnarray}
M_{i+j} &=&\sqrt{2\pi \Delta S/3}\exp (i\bar{S}+i\pi /4)  \notag
\label{eq:unif1} \\
&&{}\times \left\{ \bar{A}[J_{1/3}(\Delta S)+J_{-1/3}(\Delta S)]\right. 
\notag \\
&&\left. {}+\Delta A[J_{2/3}(\Delta S)-J_{-2/3}(\Delta S)]\right\} ,  
\\
\Delta S &=&(S_{i}-S_{j})/2,\qquad \bar{S}=(S_{i}+S_{j})/2,  \notag \\
\Delta A &=&(A_{i}-iA_{j})/2,\quad \bar{A}=(iA_{i}-A_{j})/2. \notag
\end{eqnarray}
The saddle-point approximation is recovered for large values of $\Delta S$,
using the asymptotic behavior 
\begin{equation}
J_{\pm \nu }(z)\sim \left( \frac{2}{\pi z}\right) ^{1/2}\cos (z\mp \nu \pi
/2-\pi /4)
\end{equation}
of the Bessel functions for large $z$. One should note that the uniform
approximation considers the collective contribution of a pair of saddle
points, instead of, as in the former case, taking them into account
individually.

In the classically forbidden region, one of the saddles must be discarded.
For that purpose, a branch of the Bessel function must be chosen in such a
way that the approximation exhibits a smooth functional behavior at the
Stokes transition \cite{Berry}, given by 
\begin{equation}
\mathrm{Re}\,S_{\mathbf{p}}(t_{i},t_{i}^{\prime },\mathbf{k}_{i})=\mathrm{Re}%
\,S_{\mathbf{p}}(t_{j},t_{j}^{\prime },\mathbf{k}_{j}).  \label{stokesline}
\end{equation}%
 Beyond the Stokes transition, 
\begin{eqnarray}
M_{i+j} &=&\sqrt{2i\Delta S/\pi }\exp (i\bar{S})  \notag \\
&&{}\times \left[ \bar{A}K_{1/3}(-i\Delta S)+i\Delta AK_{2/3}(-i\Delta S)%
\right] .\quad  \label{eq:unif2}
\end{eqnarray}%
The saddle-point approximation is, again, recovered using the asymptotic
expansion 
\begin{equation}
K_{\nu }(z)\sim \left( \frac{\pi }{2z}\right) ^{1/2}\exp (-z)
\label{asympt2}
\end{equation}%
for large $z.$ Inserting Eq. (\ref{asympt2}) into (\ref{eq:unif2}), it is
easy to show that only one saddle contributes to the saddle-point
approximation in this energy region. Equations (\ref{eq:unif1}) and (\ref%
{eq:unif2}) should be matched at the Stokes transitions, whose energy
positions roughly coincide with the boundary
between the classically allowed and forbidden energy regions.

In the following, we take the few-cycle pulse $\mathbf{E}(t)=-d\mathbf{A}%
(t)/dt$, with 
\begin{equation}
\mathbf{A}(t)=A_{0}\exp [-4(\omega t-\pi n)^{2}/(\pi n)^{2}]\sin [\omega
t+\phi ]\hat{e}_{x},  \label{pulse}
\end{equation}%
where $n,$ $\omega ,$ $A_{0}$ and $\phi $ denote its number of cycles,
frequency, amplitude and absolute phase, respectively. We then find the
start and return times such that the saddle-point equations are fulfilled,
and use such times to compute the yields, which are given by%
\begin{equation}
\Gamma (p_{1\parallel },p_{2\parallel })=\int d^{2}p_{1\perp }\int
d^{2}p_{2\perp }|M|^{2},  \label{yield}
\end{equation}%
where $M$ is given by Eq. (\ref{prescatt}) within the uniform approximation.

\section{Quantum-orbit analysis}

\label{quantumorb}

In Fig. 1 we present the momentum distributions computed using the
above-discussed method, for various absolute phases, in form of contour
plots in the $\left( p_{1\parallel },p_{2\parallel }\right) $ plane. We
choose the atomic species to be neon, for which electron-impact ionization
is the dominant physical mechanism \cite{footnnear}. In general, such
distributions exhibit circular shapes, centered at particular momenta along $%
p_{1\parallel }=p_{2\parallel }=p_{\parallel },$ and, in contrast to the
monochromatic-field case, are no longer symmetric with respect to $%
(p_{1\parallel },p_{2\parallel })$ $\leftrightarrow (-p_{1\parallel
},-p_{2\parallel })$. This symmetry breaking is expected, since the relation 
$A(t)=-A(t\pm T/2),$ and thus $|M(t,t^{\prime},p_{1\parallel },p_{2\parallel
})|=|M(t\pm T/2,t^{\prime}\pm T/2,-p_{1\parallel },-p_{2\parallel })|$, which
was true for monochromatic driving fields, does not hold. The circular
shapes are typical for the contact-type interaction, and are also observed
in the monochromatic case.

\begin{figure}[tbp]
\begin{center}
\includegraphics[width=8cm]{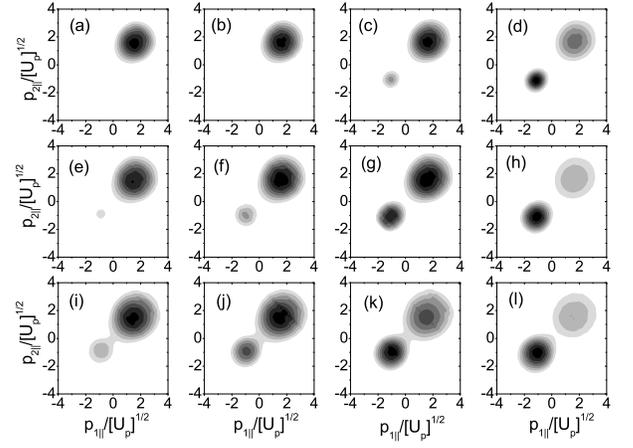}
\end{center}
\caption{Differential electron momentum distributions computed for neon ($%
|E_{01}|=0.79$ $\mathrm{a.u.}$ and $|E_{02}|=1.51$ $\mathrm{a.u.)}$ subject
to a four-cycle pulse $(n=4)$ of freqency $\protect\omega =0.057$ $\mathrm{%
a.u}$. and various intensities and absolute phases. The upper, middle and
lower panels correspond to $I=4\times 10^{14}\mathrm{W/cm}^{2}(U_{p}=0.879\ 
\mathrm{a.u}),$ $I=5.5\times 10^{14}\mathrm{W/cm}^{2}(U_{p}=1.2$ $\mathrm{a.u%
}),$ and $I=8\times 10^{14}\mathrm{W/cm}^{2}(U_{p}=1.758$ $\mathrm{a.u}),$
respectively. The absolute phases are given as follows: Panels (a), (e) and
(i): $\protect\phi =0.8\protect\pi $; panels (b), (f) and (j): $\protect\phi %
=0.9\protect\pi ;$ panels (c), (g) and (k): $\protect\phi =1\protect\pi ;$
and panels (d), (h) and (l): $\protect\phi =1.1\protect\pi .$ }
\end{figure}
Depending on the phase, the yields are mainly concentrated either in the
regions of positive or negative parallel momenta. For instance, in the
figure, initially, the parallel momenta of both electrons are essentially
positive (Figs. 1.(a), 1.(e) and 1.(i)). As the phase increases,
contributions from negative momenta are also present, becoming more and more
significant, until the distributions are almost entirely shifted from the 
positive to the negative momentum region  (c.f. Figs. 1.(d), 1.(h) and 1.(l)). This process occurs for different intervals of
absolute phases, depending on the peak intensity of the driving field. For
the specific example presented, the higher the intensity is, the earlier the
momenta start to change sign.
\begin{figure}[tbp]
\begin{center}
\includegraphics[width=8cm]{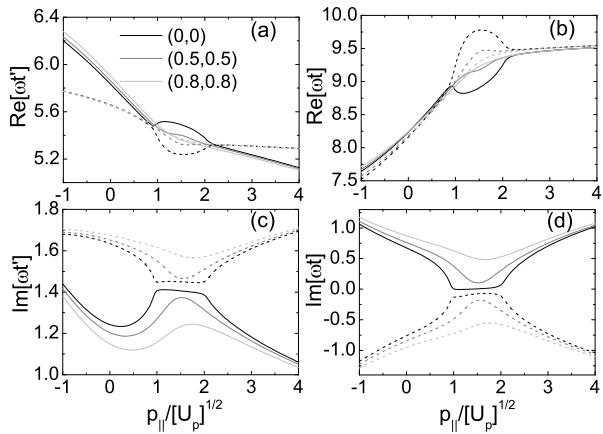}
\end{center}
\caption{Real and imaginary parts of the start and return times for the
orbits 1 and 2 as functions of the parallel momentum $p_{\parallel }$
along the diagonal $p_{1\parallel }=$ $p_{2\parallel }$, computed
for a four-cycle pulse $(n=4)$ of absolute phase $\phi =0.5\protect%
\pi $. The atomic parameters were taken as $|E_{01}|=0.79$ $\mathrm{a.u.}$
and $|E_{02}|=1.51$ $\mathrm{a.u.}$ and correspond to neon, while the field
intensity and frequency were chosen as $I=5.5\times 10^{14}\mathrm{W/cm}%
^{2}(U_{p}=1.2$ $\mathrm{a.u})$ and $\omega =0.057$ $\mathrm{a.u}$,
respectively. The numbers in the figure denote the transverse momenta $%
(p_{1\perp },p_{2\perp }$) in units of $\sqrt{U_{p}}.$ The shorter and longer orbits in each pair correspond to the  solid and dashed lines, respectively.}
\end{figure}
\begin{figure}[tbp]
\begin{center}
\includegraphics[width=8cm]{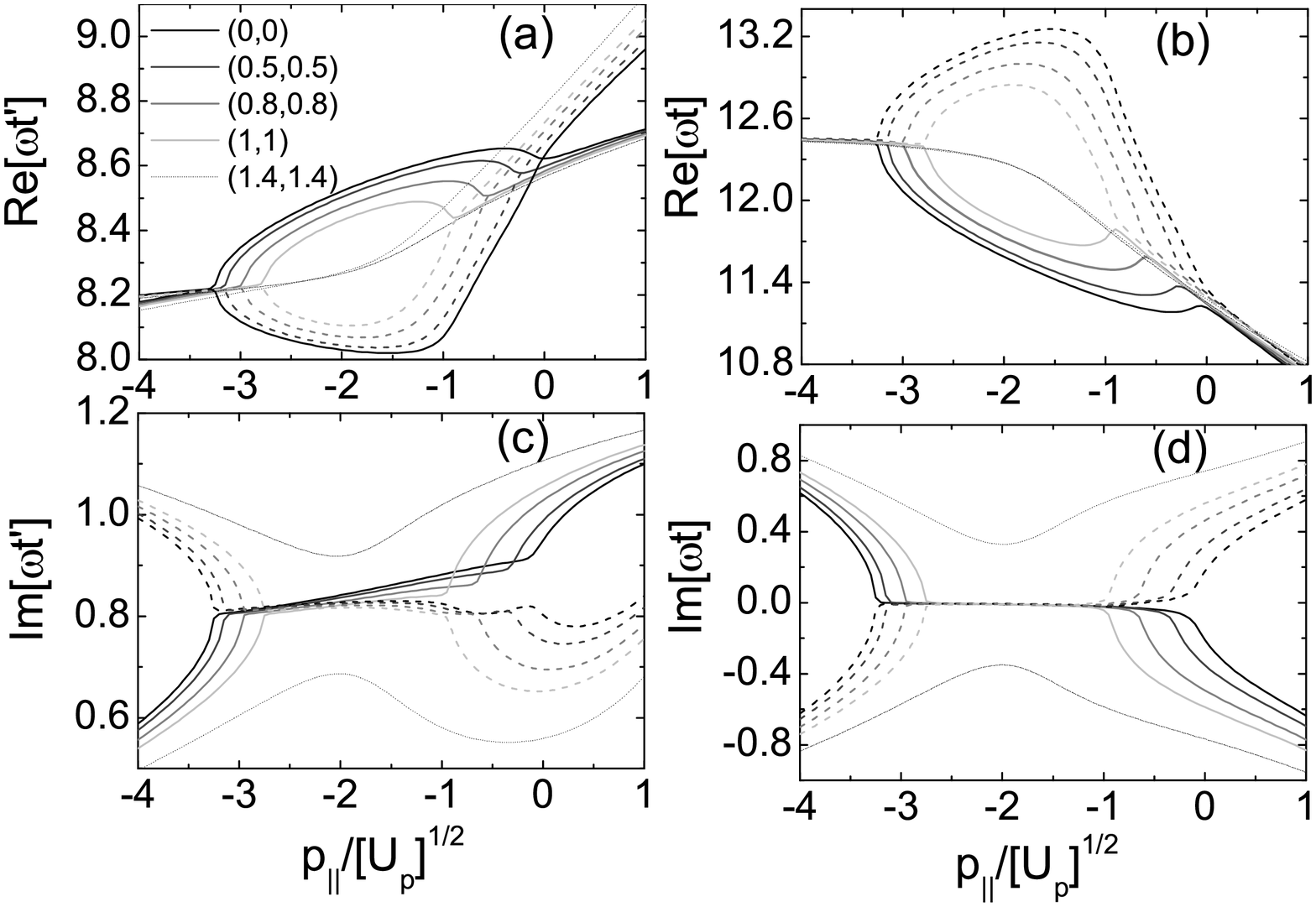}
\end{center}
\caption{The same as in the previous figures for the orbits 3 and 4.}
\end{figure}
This phase dependence is very similar to that in \cite{fewcycleclass},
obtained within a classical framework. Thereby, this behavior was traced 
back to sets of electron trajectories, whose relevance was 
determined by the phase space, and by the rate with which the first electron was 
ejected in the continuum. A critical phase was related to a change in the 
dominant pair of orbits, which had a huge repercussion in the distributions. 
This phase was also shifted towards smaller absolute 
values with increasing driving-field intensity.

Subsequently, we analyze both the asymmetry and the critical phase in
terms of pairs of quantum orbits, which are classified as $(i,j)_{(\phi )}$
according to increasing start times and absolute phases. We consider only 
relatively short orbits so
that $t-t^{\prime }\lesssim T$, where $T=2\pi /\omega $ denotes the field
cycle. Longer orbits yield negligible contributions due to wave-packet
spreading.

Fig. 2 shows one of such pairs for $t^{\prime }$ near $T$, which we
specifically denote $(1,2)_{(0.5\pi )}$. We consider the intermediate
intensity on Fig. 1, for which the phase chosen yields positive parallel
momenta, along the diagonal $p_{1\parallel }=p_{2\parallel }=p_{\parallel }$%
. Panels (a) and (b) display $\mathrm{Re}[\omega t^{\prime }]$ and $\mathrm{Re}%
[\omega t]$ as functions of $p_{\parallel }$, which can be associated to the
times obtained by solving the classical equations of motion of two electrons colliding inelastically 
in a laser field. There exists always a longer and a shorter orbit for the
electron, which nearly coalesce near two particular momenta. Such momenta correspond to Stokes transitions [Eq. (\ref{stokesline})], which, for high enough intensities, roughly coincide with the minimal and maximal classically allowed momenta \cite{footn12}. These two specific momenta delimit a region, which is most
extense for $\mathbf{p}_{j\perp }=0$ $(j=1,2)$. As the transverse momenta
increase, the effective second ionization potential $|\tilde{E}_{02}|$ also
becomes larger until this region collapses. An interesting
feature is that, between the Stokes transitions, the real parts of
the rescattering and start times are centered around a particular value of $%
p_{\parallel },$ which correspond to the peak of the momentum distributions.
For few-cycle pulses, this center depends on the pair of orbits, as well as
on the absolute phase. For monochromatic driving fields, it lies at $%
p_{\parallel }=\pm 2\sqrt{U_{p}}$ \cite{nsdiuni}.

The remaining panels depict the imaginary parts of such times, which provide
 in some sense a measure for a process being classically allowed or forbidden.
Indeed, they determine whether the transition amplitudes (\ref{prescatt})
increase or decrease exponentially, or how relevant the contributions from
particular sets of orbits are. These imaginary parts noticeably increase at
and beyond the Stokes transitions, and
remain practically constant in the momentum region inbetween. This suggests that, in this region, electron-impact ionization is either classically allowed, or at least much more probable to occur. Additionaly, whereas $\mathrm{Im}[\omega
t]$ almost vanishes in this region, $\mathrm{Im}[\omega t^{\prime }]$ has a
nearly constant and nonvanishing value (c.f. \ panels (c)). This is due to
the fact that tunneling {\it per se} is classically forbidden. Indeed, 
the larger this value is, the smaller is the probability that this process 
takes place at all.

Two additional pairs of orbits, for which $\ 1.5T\lesssim t^{\prime
}\lesssim 2T,$ are displayed in Figs. 3 and 4. At these times, the pulse (%
\ref{pulse}) is closer to its peak intensity. In such figures, there exist 
extense regions between the Stokes transitions, in which 
$\mathrm{Re}[\omega t^{\prime }]$ and $\mathrm{Re}[\omega t]$ practically 
coincide with the start and return times obtained within a classical 
framework, and in which   
$|\mathrm{Im}[\omega t]|$ are vanishingly small. Such features are 
a clear evidence that, in this case, 
electron-impact ionization is classically allowed.

Another noteworthy feature is that, in the classically allowed region, 
$|\mathrm{Im}[\omega t^{\prime }]|$ has much smaller values than those 
in Fig. 2. Physically, this means that the first
electron left with a
larger tunneling probability at $t^{\prime }$, in comparison to the orbits $%
(1.2)_{(0.5\pi )}$. Furthermore, the fact that this region is  extense shows 
that that the kinetic energy of the first electron upon return is larger for 
$(3,4)_{(0.5\pi )}$ and $(5,6)_{(0.5\pi )}$ than for $(1.2)_{(0.5\pi )}$. 
For that reason, the contributions from $(3,4)_{(0.5\pi )}$ and 
$(5,6)_{(0.5\pi )}$ to the total yield should be more relevant than those 
from  $(1.2)_{(0.5\pi )}$.

\begin{figure}[tbp]
\begin{center}
\includegraphics[width=8cm]{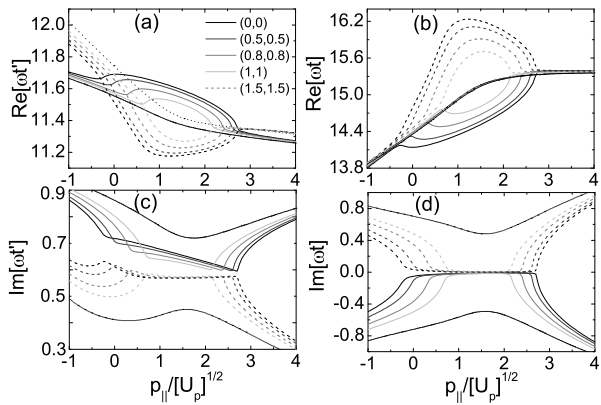}
\end{center}
\caption{The same as in the previous figures for the orbits 5 and 6.}
\end{figure}
\begin{figure}[tbp]
\begin{center}
\includegraphics[width=6cm]{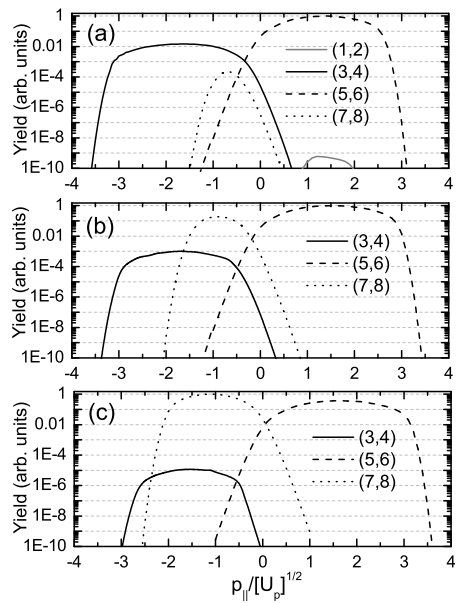}
\end{center}
\caption{Individual contributions of the four most relevant pairs of orbits to 
the NSDI yield, for $p_{1\parallel}=p_{2\parallel}=p_{\parallel}$, for ponderomotive energy  $U_p=1.2$ a.u and absolute phases  $\phi=0.5\pi$,  $\phi=0.8\pi$ and  $\phi=1.1\pi$ (panels (a), (b) and (c), respectively). The 
remaining parameters are the same as in the previous figures. The curves have been normalized to the maximum of the most relevant contributions. Specifically, in panels (b) and (c), the contributions from $(1,2)$ are smaller than the range of orders of magnitude displayed. }
\end{figure}

This is confirmed by Fig. 5, which depicts the yields computed from each
pair of the above-discussed orbits, along the diagonal $p_{1\parallel
}=p_{2\parallel }=p_{\parallel }$. In Fig. 5(a), the contributions from $%
(5,6)_{(0.5\pi )}$ are at least two orders of magnitude larger than those 
from the remaining pairs, so that the distributions will be concentrated 
in the first quadrant of the $\left( p_{1\parallel},p_{2\parallel }\right) $ 
plane. Hence, for practical purposes, the remaining contributions can be 
neglected. They are, however, very useful for the physical understanding of 
the problem.
 
The second most prominent contributions come from 
$(3,4)_{(0.5\pi )}$. This is expected, since, for these orbits, 
there is a relatively large probablity 
that the first electron tunnels out, as well as a large momentum region for 
which electron-impact ionization is allowed.

Additional contributions come from the orbits $(1,2)_{(0.5\pi )}$ 
and $(7,8)_{(0.5\pi )}$.  The latter set of orbits is not displayed in the 
previous figures, due to the fact that, in this case, there are no Stokes transitions, i.e., electron impact 
ionization is forbidden throughout.
Interestingly, the contributions from all pairs of orbits discussed above, including 
$(7,8)_{(0.5\pi )}$, are several orders of magnitude larger than those 
from the pair $(1,2)_{(0.5\pi )}$. 
This  is due to the fact that, 
for $(7,8)_{(0.5\pi )}$ the tunneling probability for the first electron is 
considerably larger than for $(1,2)_{(0.5\pi )}$. From the technical 
viewpoint, it is worth mentioning that, for $(7,8)_{(0.5\pi )}$, 
the yield has been computed by 
using Eq. (\ref{sadresca}), and taking the orbit for which this expression 
is exponentially decaying.

 For other absolute phases, there may be
other sets of orbits whose contributions may compete with or even overwhelm
those from $(5,6)$. This is in fact the case in Figs. 5(b), and 5(c), 
for $\phi=0.8\pi$ and $\phi=1.1\pi$, respectively. Such phases, as well as 
the remaining parameters, are the same as in Figs. 1(d) and 1(f), 
corresponding to the beginning and to the end of a shift in the 
momentum distributions. 

In Fig. 5(b), 
one clearly sees that the second most relevant pair of orbits is no 
longer $(3,4)_{(0.8\pi )}$, but $(7,8)_{(0.8\pi )}$. The contributions 
from such orbits now are only one order of magnitude smaller than  
those from $(5,6)_{(0.8\pi )}$. Consequently, there are also small, but
not negligible contributions in negative momentum regions. This is in 
agreement with Fig. 1(d), for which there is a small spot in the third 
quadrant of the $(p_{1\parallel}, p_{2\parallel})$ plane, in addition to 
the dominant contributions in the first quadrant. 
For larger phases, as for instance $\phi=1.1\pi$ (Fig. 5(c)), 
the contributions from  $(7,8)$ become even more relevant than those from 
$(5,6)$. Hence, the distributions are shifted from the first to the third 
quadrant, in accordance to Fig. 1(h).   
 
\begin{figure}[tbp]
\begin{center}
\includegraphics[width=6cm]{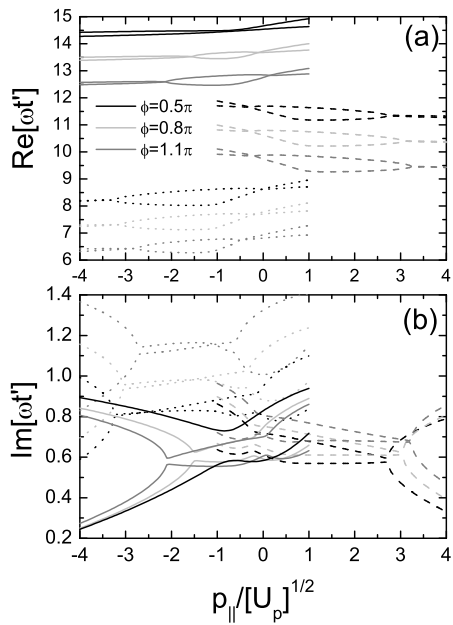}
\end{center}
\caption{Real (panel (a)) and imaginary (panel (b)) parts of the start times
for the orbits (3,4), (5,6) and (7,8) along $p_{1\parallel }=$ $%
p_{2\parallel }$ and for vanishing transverse momenta, indicated  by dotted,
dashed and solid lines, respectively. We consider neon
subject to a four-cycle \ $(n=4)$ pulse of intensity $I=5.5\times 10^{14}%
\mathrm{W/cm}^{2},$ various absolute phases and frequency $\protect\omega %
=0.057\ \mathrm{a.u}.$ Specifically for $\phi=0.5\pi$, electron-impact 
ionization is classically forbidden for the orbits $(7,8)$.}
\end{figure}
In Fig. 6, we systematically analyze the dependence of the real and
imaginary parts of the times $t^{\prime }$ on the absolute phase, for the
three most relevant sets of orbits. As an overall feature, for the pair $%
(3,4),$ $|\mathrm{Im}[\omega t^{\prime }]|$ is larger than for the remaining
two pairs. Physically, this means that tunneling is less probable in this
case. In general, this makes the contributions from this pair to the yield 
less relevant than those from the other two sets. An exception, however, 
occurs for $\phi=0.5\pi$, due to the fact that, in this case, electron-impact ionization is forbidden for $(7,8)$. This leads to exponentially decaying contributions for this pair, which are overwhelmed by those from $(3,4)$. Still, even in this particular case, these latter are two order of magnitude smaller than the
yield from the orbits $(5,6)$ (c.f. Fig. 5(a)).

For high driving-field intensities, as
those in the lower panels in Fig. 1, there may exist minor, though not
negligible, contributions from $(3,4)$. In general, however, these
contributions are vanishingly small. An interesting feature is that, with 
increasing absolute phases, this set of orbits gradually loses relevance, 
since $|\mathrm{Im}[\omega t^{\prime }]|$ increases. This can be explicitly 
seen in Fig. 5, where the yield from $(3,4)$ decreases in at least three 
orders of magnitude as $\phi$ increases.

The other two pairs of orbits, $(5,6)$ and $(7,8)$ are, in fact, far more important to the yield.
Thereby, three distinct behaviors can be identified. Below $\phi =0.8\pi ,$
the distributions are mainly determined by the orbits $(5,6),$ whose
contributions lie in the region of $p_{\parallel }>0$. Around this phase, 
the pair $(7,8)_{(0.8\pi )}$ comes into play. 
Indeed, although this pair does not delimit a large classically allowed
region, the imaginary parts of the corresponding start times are comparable
to or smaller than those of $(5,6).$ Thus, the contributions from both pairs
start to compete, and the distributions spread over both positive and
negative momentum regions. As the phase increases, $|\mathrm{Im}[\omega
t^{\prime }]|$ gradually decreases for this latter set, until the negative
parallel momenta are favored. This explains the features in Fig. 1.
One may refer to $\phi=0.8\pi$ as a critical phase $\phi_c$, since it marks 
a change in the sets of dominant orbits.

\begin{figure}[tbp]
\begin{center}
\includegraphics[width=6cm]{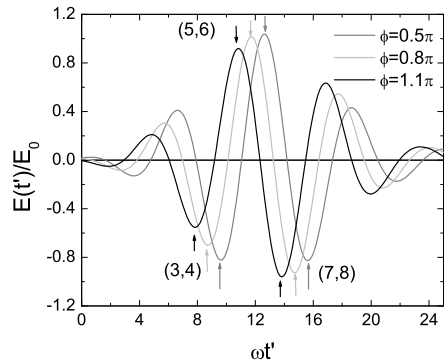}
\end{center}
\caption{Time-dependent electric field, for various absolute phases. The
peak-field times are marked with arrows and the corresponding pairs of
orbits are indicated by the numbers in the figure. 
The remaining parameters are the same as in the previous figure.}
\end{figure}
\begin{figure}[tbp]
\begin{center}
\includegraphics[width=6cm]{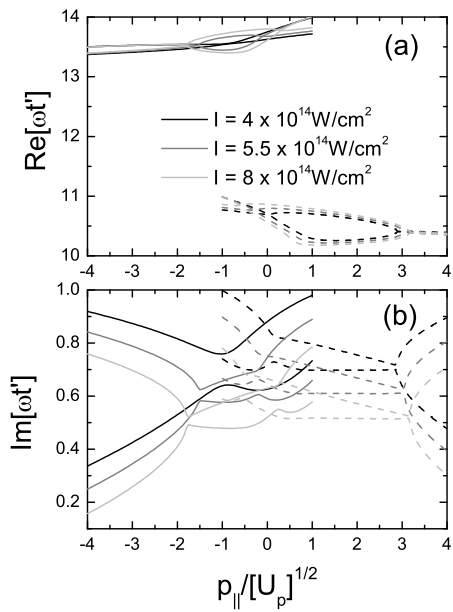}
\end{center}
\caption{Real (panel (a)) and imaginary (panel (b)) parts of the start times
for the orbits (5,6), (7,8), indicated by dashed and solid lines,
respectively, for absolute phase $\protect\phi =0.8\protect\pi $ and various
laser intensities, along $p_{1\parallel }=$ $p_{2\parallel }$ and vanishing
transverse momenta. The remaining parameters are the same as in the previous
figure.}
\end{figure}
The pulse profile (Fig. 7), together with the real parts of the tunneling
times, allows an intuitive interpretation of the above statements. Below $%
\phi _{c}=0.8\pi ,$ the peak value of the pulse is near $\mathrm{Re}[\omega
t^{\prime }]$ for the orbits $(5,6),$ whose contributions then dominate.
Thus, the distributions essentially concentrate in the positive momentum region.
Around this phase, this picture starts to change, and there are two sets of
orbits, namely $(5,6)$ and $(7,8),$ for which the instantaneous electric
fields at the tunneling times are comparable. This situation persists within
a phase interval until, finally, \ the absolute maximum of the field
corresponds to the latter set of orbits, so that the momenta are mainly
negative. An interesting situation occurs for $\phi =0.5\pi $, $\ $for
which, in principle, there exists two sets of times near which the electric
field exhibits comparable maxima, corresponding to $(3,4)$ and $(7,8)$. For
low intensities, rescattering is not allowed for the orbits $(7,8)$, and the
contributions to the yield are absent. For high intensities, however,
electron-impact ionization is already allowed in this case, so that the
critical phase is shifted towards smaller values.

In Fig. 8, we explicitly show how electron-impact ionization becomes 
classically allowed or  or forbidden
upon a change in the driving-field intensity, for $\phi _{c}=0.8\pi $. In
fact, as the intensity is decreased, this process ceases to be
allowed for the orbits $(7,8),$ whose contributions lie near \ the peak of
the pulse (c.f. Fig. 7). As a direct consequence, the critical phase is
shifted towards larger absolute values as the intensity decreases, as shown
in Fig. 1.

\section{Comparison with classical models}

\label{classical}

In this section, we will recall and apply the classical model used in \cite%
{fewcycleclass}, performing a direct comparison with the results of the
$S$-Matrix computation. We consider an electron ensemble subject to the
few-cycle pulse (\ref{pulse}), which are released from the origin of the
coordinate system with initial vanishing drift velocities, i.e.,%
\begin{equation}
\mathbf{p}+\mathbf{A}(t^{\prime })=0.  \label{drift}
\end{equation}%
The varying parameters are the tunneling times $t^{\prime },$ uniformly
distributed throughout the pulse, and the quasi-static tunneling rate \cite%
{quasistatic} 
\begin{equation}
R(t^{\prime })\sim \exp \left[ -2(2|E_{01}|)^{3/2}/3|E(t^{\prime })|\right]
/|E(t^{\prime })|,  \label{quasistat}
\end{equation}%
with which the electron counts are weighted. Some of these electrons
subsequently return to the origin and free a second electron ensemble at
later times $t$ through electron-impact ionization. Their return and
rescattering conditions are given by Eqs. (\ref{saddle3}) and (\ref{saddle2}%
), respectively. The yields are then given by

\begin{equation}
\Gamma \sim \int dt^{\prime}R(t^{\prime })\delta \left( E_{\mathrm{ret}%
}(t)-\sum\limits_{j=1}^{2}\frac{(\mathbf{p}_{j}+\mathbf{A}(t))^{2}}{2}%
-|E_{02}|\right)  \label{classicalyield}
\end{equation}%
where $E_{\mathrm{ret}}(t)$ is the kinetic energy of the electron upon
return and the argument in the $\delta $ function gives the energy
conservation at $t$. This model is discussed in more detail in \cite{preprint2}%
.

Fig. 9 presents the outcome of the classical computation, for the same
parameters as in Fig. 1. Interestingly, both figures are very similar.
Indeed, there exist only minor differences near the boundaries of the
classically allowed region, which occur for high intensities, 
as displayed in the lower panels of Figs. 1 and 9. Such 
differences are
due to the fact that the yield from the quantum-mechanical computation 
is exponentially decaying in the region for which electron-impact ionization
is  classically forbidden, whereas the distribution (\ref{classicalyield}) 
immediately vanishes. Such discrepancies were also present in the monochromatic case.

Further 
discrepancies occur for specific phases, in panels (c) and (e) of both
figures, and show that the sign reversal starts to take place for 
slightly smaller absolute phases in the quantum-mechanical case. 
This effect can be understood if one keeps in mind that the critical phase 
indicates a change in the dominant set of trajectories, and that 
pre-requisites for this dominance are a large tunneling probability for the 
first electron and electron-impact ionization being classically allowed. 
In the particular example provided in the panels (c) and (e) of 
Figs. 1 and 9, this process has just become allowed for the orbits $(7,8)$, 
within a small momentum region. In the quantum-mechanical case (Fig. 1), 
there exist contributions to the yield from near the boundary of such a region, 
even if rescattering is forbidden, whereas in the classical model (Fig. 9) 
such contributions vanish. Obviously, such discrepancies are absent in the 
symmetric momentum distributions obtained in the monochromatic case.

For 
monochromatic driving fields, a very good agreement has also been 
reported in previous publications \cite{preprint1,preprint2}. 
It is not obvious, however, that this would remain true in the few-cycle 
regime.  
In fact, apart from yielding distributions which immediately vanish at the 
boundaries of the classically allowed region, the classical model does not 
take into account several effects 
which are present in the quantum-mechanical computation. Examples of such 
effects are the quantum interference between different possible paths 
for the returning electron, or the spread of the electron wave packet. 
Furthermore, the classical model considers an additional approximation 
with respect to the $S$-Matrix computation, namely Eq. (\ref%
{quasistat}), which is a quasi-static, cycle-averaged tunneling rate.
This rate is a key ingredient in the distributions given by 
Eq. (\ref{classicalyield}), and vital for 
the phase-dependence observed in Fig. 9 (c.f. Fig. 10 and discussions 
in \cite{fewcycleclass}).
However, its validity may be limited or even questionable in the few-cycle 
regime. Indeed, recently, a non-adiabatic rate that should be more accurate 
in this case has been derived \cite{rateyudivanov}. In general, the 
non-adiabatic rate tends to broaden the time range for which the 
relevance of sets of orbits persists. We verified, however, that this rate
does not modify the yields for the parameter range in question.
Discrepancies between the quasi-static and the non-adiabatic rate occur only
when the Keldysh parameter $\gamma =\sqrt{|E_{01}|/(2U_{p})}$ is much larger
than unity. In this case, the driving-field intensity would be far too low
for the rescattering process discussed in this paper to be allowed, and thus
for the classical model to be applicable. For measurements of NSDI yields
below the threshold see, e.g., Ref. \cite{belowthresh}.

\begin{figure}[tbp]
\begin{center}
\includegraphics[width=8cm]{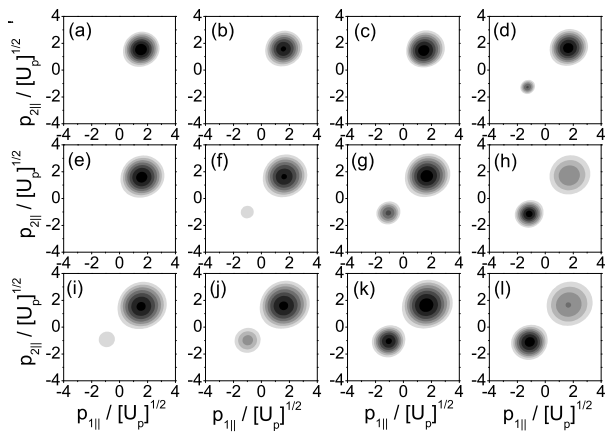}
\end{center}
\caption{Differential electron momentum distributions for the same
parameters as in Fig.1, computed with the classical model discussed in Sec.
IV. }
\end{figure}
The subsequent picture (Fig. 10) is the classical counterpart of Fig. 6,
providing an interpretation of Fig. 9 in terms of 
the interplay between the ionization rate (\ref{quasistat}) and the phase 
space. Thereby, as in the quantum-mechanical case, we consider parallel 
momenta along the diagonal $p_{1\parallel}=p_{2\parallel}=p_{\parallel}$ 
and vanishing transverse momenta. The main contributions to the yield 
come from pairs of trajectories for which the tunneling probability is large, 
and for which electron-impact ionization is classically allowed. 
In the figure, according to such criteria, 
it is  possible to identify two sets of relevant orbits, corresponding to 
electrons ejected at $9 \lesssim \omega t^{\prime} \lesssim 12$, with positive 
parallel momenta, and to electrons released at 
$12 \lesssim \omega t^{\prime}\lesssim 15$, with negative parallel momenta. 

According to the absolute phase, similarly to the quantum-mechanical case, 
there exist three distinct types of behavior. For $\phi<0.8\pi$, 
apart from the fact that the classically allowed region is almost vanishing for the 
orbits $(7,8)$, the quasi-static rate (\ref{quasistat}) is considerably larger for the former set of electrons, 
so that the distributions are 
concentrated in the positive momentum regions. Around $\phi=0.8\pi$, 
such tunneling rates are comparable for both sets of orbits, resulting in nonvanishing 
yields for positive and negative parallel momenta. Finally, as the phase 
increases, tunneling is more prominent for the latter set of orbits, and 
the distributions gradually change towards negative 
momenta. As in the previous section, the critical phase marks a 
change in the dominant orbits. This is in agreement with Fig. 6 and with the
results in \cite{fewcycleclass}.
\begin{figure}[tbp]
\begin{center}
\includegraphics[width=8cm]{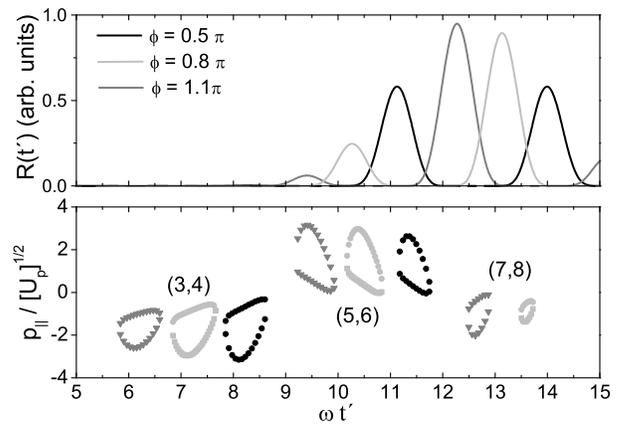}
\end{center}
\caption{Quasi-static tunneling rate (\ref{quasistat})(panel (a)), 
together with the classically allowed momenta computed with the classical 
model (panel (b)), as functions of the tunneling times, for the same 
parameters as in Fig. 6.}
\end{figure}

The ratios between the individual contributions of such pairs of orbits, as well as the momenta for which their maxima occur, displayed in Fig. 11, are also in very good agreement with its quantum-mechanical counterpart (Fig. 5). However, the distributions computed with the classical model are narrower than those obtained using the quantum-mechanical computation. This is once more related to the fact that, in a classical framework, the contributions from the forbidden momentum region can not be taken into account.

\begin{figure}[tbp]
\begin{center}
\includegraphics[width=6cm]{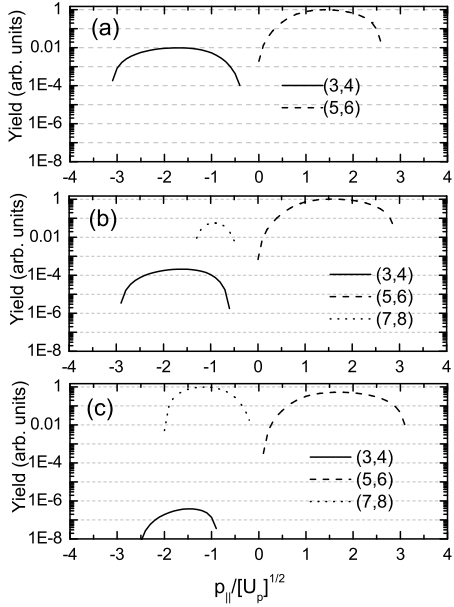}
\end{center}
\caption{Individual contributions to the momentum distributions for the same pairs of orbits and for the same  parameters as in Fig. 5, computed with the classical model. Parts (a), (b) and (c) correspond to $\phi=0.5\pi$, $\phi=0.8\pi$ and $\phi=1.1\pi$, respectively. The curves have been normalized to the peak value of the largest curve. The contributions from the orbits for which electron-impact ionization is classically forbidden are  absent.}
\end{figure}
\begin{figure}[tbp]
\begin{center}
\includegraphics[width=6cm,angle=270]{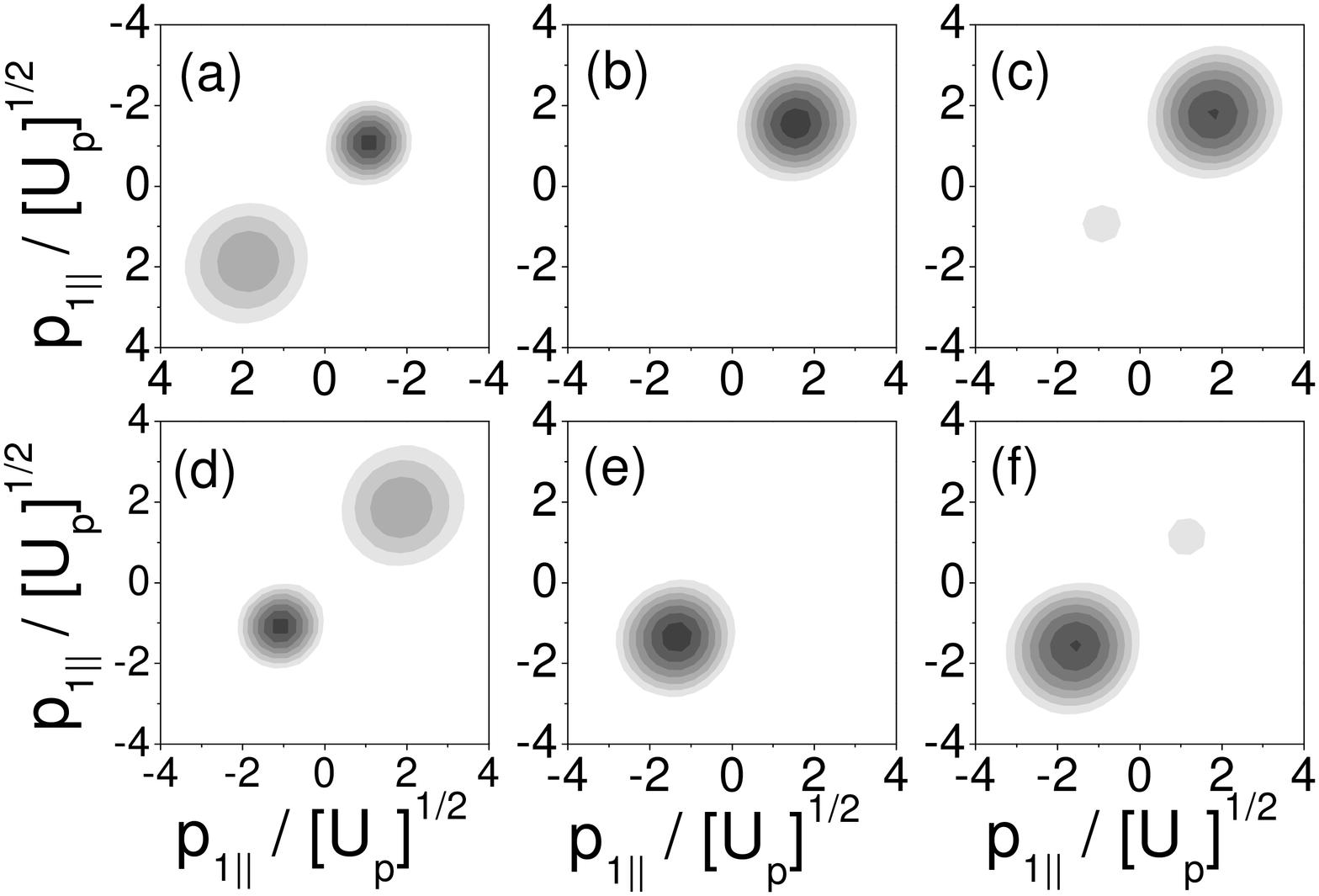}
\end{center}
\caption{Differential electron momentum distributions computed with the classical model 
in Sec. IV for neon subject to a 
four-cycle pulse $(n=4)$ of approximate intensity $I=5.5 \times 10^{14}\mathrm{W/cm}^2$ (Up=1.2), 
frequency $\omega=0.057$ a.u. and absolute phases $\phi=0.1\pi$, $\phi=0.5\pi$,
and $\phi=0.9\pi$ (panels (a) to (c), respectively), and  $\phi=1.1\pi$, $\phi=1.5\pi$,
 and $\phi=1.9\pi$ (panels (d) to (f), respectively).}
\end{figure}

Another interesting feature, which is displayed in Fig. 12, is the existence
of other critical phases. For instance, around $\phi =0.1\pi ,$ there is a
transition in the momentum distributions from the third to the first
quadrant in the $(p_{1\parallel },p_{2\parallel })$ plane, i.e., exactly in
the opposite direction as the transition in Figs. 1 and 9. Furthermore, the
upper and lower panels in Fig. 12 look exactly the same, if the first and
the third quadrant are interchanged. This shows that there is a symmetry in
the momentum distributions, which is due to the fact that $\mathbf{A}(t,\phi
)=-\mathbf{A}(t,\phi \pm \pi ),$ so that $|M(p_{1\parallel },p_{2\parallel
},\phi )|=|M(-p_{1\parallel },-p_{2\parallel },\phi \pm \pi )|.$

\section{Conclusions}

\label{concl}

The studies performed in this paper clearly show that nonsequential double
ionization with few-cycle pulses is a powerful tool for absolute-phase
measurements. More specifically, the yields, as functions of the electron
momentum components $(p_{1\parallel },p_{2\parallel })$ parallel to the
laser-field polarization, are mainly concentrated either in the positive or
negative momentum region, depending on the absolute phase in question.
Around a critical phase, such distributions start to shift from one momentum 
region to the other, until, as the phase increases, complete sign reversal 
in the momenta occurs. Such features, obtained considering $(e^{-},2e^{-})$ 
electron-impact ionization within a
quantum-mechanical $S$-Matrix framework, are interpreted in terms of the
so-called quantum orbits, which can be directly associated to the
trajectories of classical electrons. Both the asymmetry and the critical
phase result from the interplay between phase-space effects, and the
probability that the first electron leaves its parent ion through tunneling
ionization. The former and the latter, respectively, determine whether
electron-impact ionization is classically allowed or forbidden, or the
relevance of a set of orbits to the yield.

 The huge effects observed, namely
the yields vanishing or appearing over extense and well-separated regions in
the $(p_{1\parallel },p_{2\parallel })$ plane, are a due to a particular
characteristic of the rescattering process in question, for which, in
addition to a maximal, there is also a minimal classically allowed energy.
In other words, electron-impact ionization is allowed only within confined
phase-space regions, defined by the radius in Eq. (\ref{saddle4}). By varying the 
driving-field intensity and the absolute phase adequately, this radius can 
be forced to vanish, so that a whole phase-space region would become classically forbidden. 
Furthermore, a particular region can be made irrelevant due to a small 
tunneling probability for the first electron. This is a major
advantage over other phenomena occurring in the context of strong-laser
field matter interaction, such as above-threshold ionization and high-order
harmonic generation. For both phenomena, there are only maximal classically
allowed energies, so that these effects do not occur.

Apart from providing support for previous classical computations \cite%
{fewcycleclass}, the present results allow one to establish a one-to-one
correspondence between the classical and the quantum-mechanical approaches.

Tunneling, for instance, is incorporated in the classical model using the
quasi-static rate (\ref{quasistat}) which weights the electron counts. In the
$S$-Matrix computation, this process is directly related to the imaginary
parts of the start times $t^{\prime }.$ Both the quasi-static rate and  $%
\mathrm{Im}[t^{\prime }]$ are somehow a measure of the relevance of a set of
orbits. Indeed, dominant contributions always come from pairs of orbits for
which Eq.(\ref{quasistat}) is larger, or $\mathrm{Im}[t^{\prime }]$ are
smaller than those from the remaining pairs. Depending on the absolute
phase, this may occur for a single set of orbits, whose contributions lie
either in the negative or in the positive momentum regions, or there may be
sets of orbits whose contributions compete. A critical phase characterizes a
change in the dominant set of orbits.

Furthermore, the phase-space effects which occur if electron-impact 
ionization  becomes classically forbidden, i.e., if the radius in 
Eq. (\ref{saddle4}) collapses, are present in both frameworks in 
very similar, though not entirely identical, ways. In the
classical computation, this would lead to vanishing yields, 
since the condition in the argument of the \ $\delta $ function 
in Eq. (\ref{classicalyield}) would
never be fulfilled. In the quantum-mechanical model, there would be
exponentially decaying contributions throughout. If this radius does not
collapse, the process will be allowed within a confined phase-space region.
 In the classical case, the start and return times coalesce at the boundaries
of this region, whereas, quantum-mechanically, this does not completely
happen. 

The above-stated effects explain the minor 
discrepancies between Figs. 1 and 9. 
In particular, two types of discrepancies have been observed. First, for high 
driving-field intensities, the distributions obtained with the 
quantum-mechanical computation are slightly broader than those from the 
classical model. 
Second, in the quantum-mechanical framework, the distributions start 
to shift at a slightly smaller phase, as compared to the outcome of the 
classical simulation.  The former differences were also present in NSDI with 
monochromatic driving fields, whereas the latter discrepancies are specific to
asymmetric momentum distributions, which occur, for instance, in the context 
of few-cycle driving pulses.

The agreement between both the quantum-mechanical and the classical
computations go beyond the physical explanations for the asymmetry and the
critical phase. Indeed, the momentum distributions computed with the one or
the other method, as well as the predicted critical phase and the interval
in which the momenta change sign, are very similar, apart from minor 
differences near the classical boundaries. This is a
concrete evidence that the effects reported in this paper are not rooted on
a particular model or framework, being, on the contrary, of a deeper
physical nature. In fact, recently, a similar asymmetry has been observed in
ongoing NSDI experiments with few-cycle driving pulses \cite{xjliuwien}.

\acknowledgements{Discussions with W. Becker and H. Rottke are gratefully 
acknowledged. This work was supported in part by the Deutsche 
Forschungsgemeinschaft (European Graduate College ``Interference and 
Quantum Applications'' and SFB407).}


\begin{references}
\bibitem{footnoteup} The ponderomotive energy  is 
defined as the time average $U_p=<\mathbf{A}(t)^2>_t/2$, where 
$\mathbf{A}(t)$ denotes the vector potential. In this paper, in order to 
compare our results more directly with the existing literature, we take 
the expression for $U_p$ in monochromatic driving fields, i.e., $U_p=A^2_0/4$.
\bibitem{footnsfa} The strong-field approximation mainly consists in neglecting 
the binding potential in the propagation of the electron in the continuum, the laser 
field when the electron is bound, or at the rescattering, and the excited states of 
the atom in question.
\bibitem{footn12}Specifically for the orbits $(1,2)$, if the first ionization potential vanishes, i.e., in the classical limit, electron-impact ionization 
is only classically allowed for slightly larger intensities, corresponding 
to $U_p=1.218$ a.u., whereas, in the $S$-Matrix framework, there exist Stokes 
transitions already for $U_p=1.2$ a.u. This is related to the fact that, for
$(1,2)$, the driving-field intensity at the tunneling time $t^{\prime}$ is 
low enough for the quantum-classical correspondence to become less reliable.
This behavior shows that the existence of Stokes transitions is a necessary, but not sufficient condition for the existence of an energy region for which electron-impact ionization is classically allowed.
\end{references}
\end{document}